ARTICLE

# Structure evolution of nanoparticulate $Fe_2O_3$

Andreas Erlebach, Heinz-Dieter Kurland, Janet Grabow, Frank A. Müller and Marek Sierka*

Atomic structure and properties of nanoparticulate $Fe_2O_3$ are characterized starting from its smallest $Fe_2O_3$ building unit and $(Fe_2O_3)_n$ clusters through to nanometer-sized $Fe_2O_3$ particles. This is achieved combining global structure optimizations at the density functional theory level, molecular dynamics simulations employing tailored, *ab initio* parameterized interatomic potential functions and experiments. With exception of nearly tetrahedral, adamantane-like $(Fe_2O_3)_2$ small $(Fe_2O_3)_n$ clusters assume compact, virtually amorphous structures with little or no symmetry. For $n$ = 2–5 $(Fe_2O_3)_n$ clusters consist mainly of two and three-membered Fe–O rings. Starting from $n$ = 5 they increasingly assume tetrahedral shape with adamantane-like $(Fe_2O_3)_2$ unit as the main building block. However, the small energy differences between different isomers of the same cluster size make precise structural assignment for larger $(Fe_2O_3)_n$ clusters difficult. The tetrahedral morphology persists for $Fe_2O_3$ nanoparticles with up to 3 nm in diameter. Simulated crystallization of larger nanoparticles with diameters of about 5 nm demonstrates pronounced melting point depression and leads to formation of ε-$Fe_2O_3$ single crystals with hexagonal morphology. This finding is in excellent agreement with the results obtained for $Fe_2O_3$ nanopowders generated by laser vaporization and provides the first direct indication that ε-$Fe_2O_3$ may be thermodynamically the most stable phase in this size regime.

## Introduction

One of the fundamental questions of nanoscience is how the structure and properties of a material change with increasing aggregation state, starting from individual atoms, small clusters and nanoparticles through to bulk material. Here we answer this question for one of the technologically very important[1] materials – iron(III) oxide. Nanoparticulate $Fe_2O_3$ finds a number of applications due to its unique magnetic, biochemical and catalytic properties at the nanoscale.[2] It is used in catalysis, biomedicine for hyperthermia based anticancer therapy and targeted drug delivery, magnetic resonance imaging and immunoassays as well as for magnetic data storage.[2-4] $Fe_2O_3$ shows pronounced size dependence of its structural and magnetic properties.[5,6] As bulk material it exists either as the thermodynamically most stable hematite (α-$Fe_2O_3$) or metastable maghemite (γ-$Fe_2O_3$), both naturally occurring as minerals. In addition, two $Fe_2O_3$ polymorphs are known, β-$Fe_2O_3$ and ε-$Fe_2O_3$ that can only be obtained in nanoparticulate form.[5] ε-$Fe_2O_3$ has attracted particular attention since the discovery of its unique magnetic and dielectric properties[6] such as the giant coercive field due to the large magneto-crystalline anisotropy and a relatively small saturation magnetization.[7] This makes ε-$Fe_2O_3$ a promising candidate for advanced materials. Its crystal structure can be described as intermediate between α- and γ-$Fe_2O_3$, containing one-quarter of Fe atoms in tetrahedral interstices and three-quarters in octahedral sites. A unique feature ε-$Fe_2O_3$ is the presence of five-fold coordinated O atoms. It has been suggested[8] that below certain particle size ε-$Fe_2O_3$ may be thermodynamically the most stable phase. However, this assumption has not yet been explicitly verified.[6] It addition to the crystalline forms it is also possible to obtain amorphous $Fe_2O_3$.[5]

Among different synthetic routes gas phase processes, such as flame spray pyrolysis,[9] laser ablation,[10] plasma synthesis[11] and the laser vaporization of iron oxide raw powders[12] are convenient fabrication methods of $Fe_2O_3$ nanoparticles (NP). Due to the size dependence of structure and properties of $Fe_2O_3$ NP a greater understanding of the nucleation, growth and crystallization mechanisms is essential to control and optimize their synthesis. Apart from a fundamental interest small $Fe_2O_3$ clusters are also important intermediates in the initial gas-phase nucleation stages.[13] The elucidation of their atomic structure and properties is a crucial step towards detailed understanding of NP formation processes. In addition, nanoclusters can display chemical and physical properties distinct from both small molecules and the corresponding bulk materials or larger NP.[14] For example, magnetic properties of small iron oxide clusters and $Fe_2O_3$ NP with diameters between 1-5 nm are strongly size dependent.[15] The knowledge of atomic structure in this size range is one of the key prerequisites for determination and control of such structure-property relationships. However, atomic level characterization of iron oxide nanoclusters is a very challenging task due to its complicated electronic structure. In addition, computational studies of larger clusters are generally hampered by a steeply increasing number of local minima with increasing cluster size.[13,16] This makes the search for low energy structures by manual construction of all possible isomers followed by local structure optimizations very challenging. Therefore, several global energy minimization techniques for automatic determination of the most stable cluster structure have been proposed.[17,18] Among them, genetic algorithm (GA) finds the global minimum structure by an evolutionary process[13,18,19] and has been applied successfully for structure predictions of various metal oxide nanoclusters (see, *e.g.*, Ref. 16).

Due to these difficulties computational studies of iron oxide clusters have so far been limited to small non-stoichiometric ionic[20-26] and neutral species.[27,28] Stoichiometric, neutral $(Fe_2O_3)_n$ clusters were investigated up to $n = 2$[29-31] as well as for $n = 2–6$ and $10$[32] using only manually constructed and locally optimized structures. Recently, we reported the first global structure optimization of $(Fe_2O_3)_n$ clusters with $n = 1–5$ employing density functional theory (DFT) and including precise determination of their magnetic (spin) states.[33] This study demonstrated that the geometric structure of larger $(Fe_2O_3)_n$ clusters is virtually independent of their magnetic configurations. In addition, starting from $n = 4$ the precise spin state has only a minor influence on relative energies of different cluster isomers.

The steep increase of computational cost of global structure optimization algorithms makes search for global energy minima of larger clusters and nanoparticles virtually impossible. Even if the global minimum of a large system could eventually be located the large number of very close-lying local minima renders the result meaningless. Instead, low-energy structures can be located employing molecular dynamics (MD) simulations along with carefully parameterized interatomic potential functions (IP).[13,34] As an example, simulated annealing procedure was employed to investigate crystallization process and structure of metal oxide NP with several nm in diameter.[35-37] However, similar studies of $Fe_2O_3$ NP with diameters in the range of 2-5 nm lead only to amorphous structures,[38] most probably due to short simulation times and shortcomings of the potential functions employed. Despite the unique properties, technological relevance and complex polymorphous transformations of crystalline $(Fe_2O_3)_n$ NP no computational investigations of their structure, formation and crystallization processes have been reported so far.

Our present work is the first systematic, comprehensive study of nanoparticulate $Fe_2O_3$ starting from its smallest $Fe_2O_3$ building unit and $(Fe_2O_3)_n$ clusters of increasing size through to nanometer-sized $Fe_2O_3$ particles. The key ingredient are tailored, *ab initio* parameterized interatomic potential functions (IP-$Fe_2O_3$). Combined with a refinement at the DFT level they are used to locate global minima of $(Fe_2O_3)_n$ clusters with $n = 1$-$10$. Finally, the IP-$Fe_2O_3$ are used for simulated crystallization of $Fe_2O_3$ NP with diameters up to 5 nm. The results of the simulations of such large NP are in excellent agreement with results obtained for $Fe_2O_3$ NP generated by the laser vaporization (LAVA)[39] process.

## Computational details

*Ab initio* derived interatomic potentials are the main computational tool used in the present study. Their functional form $\phi_{ij}$ is based on the Born-Mayer (BM) model for ionic solids[40] with

$$\phi_{ij}(r_{ij}) = \frac{q_i q_j}{r_{ij}} + A_{ij} exp\left(-\frac{r_{ij}}{\rho_{ij}}\right) - \frac{C_{ij}}{r_{ij}^6}, \quad (1)$$

where $r_{ij}$ is the interatomic distance between the centers $i$ and $j$ with charges $q_i$ and $q_j$. The second and third term in Eq. (1) with adjustable parameters $A_{ij}$, $\rho_{ij}$ and $C_{ij}$ describe short-range repulsive and attractive interactions, respectively. These parameters for Fe–O, O–O and Fe–Fe interactions along with the corresponding charges were derived employing a least square fitting procedure as implemented in the GULP program.[41-43] The training data set included DFT calculated structures and energies of bulk $Fe_2O_3$ polymorphs: α-$Fe_2O_3$ (hematite, orthogonal unit cell),[44,45] γ-$Fe_2O_3$ (maghemite)[46] and ε-$Fe_2O_3$.[6,47] In addition, β-$Fe_2O_3$[48,49] served as a test for the transferability of the potentials.

All DFT calculations for the bulk structures were performed employing periodic boundary condition and the Vienna *ab initio* simulation package (VASP),[50,51] the Perdew-Burke-Ernzerhof (PBE)[52,53] exchange-correlation functional along with the projector augmented wave (PAW) method.[54,55] In order to improve the description of Fe 3d states a Hubbard correction was included employing the DFT+U methodology[56-58] in the simplified, rotationally invariant Dudarev's form[59] with an effective parameter $U_{eff} = 4$ eV. This approach mimics the effects of on-site coulomb and exchange interaction between electrons with considerably less computational demands compared to the computation of exact exchange within hybrid exchange-correlation functionals. Therefore, DFT+U methodology was successfully applied for simulations of structural, electronic and magnetic properties of bulk $Fe_2O_3$ in a good agreement with experimental data.[46,60] The empirical van der Waals correction of Grimme *et al.* was also added.[61] Plane wave basis sets used 600 eV cutoff and integration of the first Brillouin zone employed 6×8×4 (α-$Fe_2O_3$), 6×6×6 (β-$Fe_2O_3$), 6×6×2 (γ-$Fe_2O_3$) and 8×6×6 (ε-$Fe_2O_3$) Monkhorst-Pack[62] grids.

The bulk structures of α-, γ- and ε-$Fe_2O_3$ were first optimized assuming the magnetic ground state (GS) in each case. Next, single point calculations for isotropic and anisotropic unit cell deformations were performed. The final training set parameterization of interatomic potentials included 10 isotropic distortions with volume changes between -10 and +10% as well as 24 anisotropic distortions for each $Fe_2O_3$ polymorph. The anisotropic distortions were created by expansion and compression of individual unit cell vectors, with the variations of ±5, ±2.5 and ±1.7% for the simultaneous change of one, two or three vectors, respectively. The bulk moduli of α- and γ-$Fe_2O_3$ used to test the quality of the potential were fitted to the Birch-Murnaghan equation of state[63] using energies of the isotropic deformed unit cells computed at the IP-$Fe_2O_3$ and DFT level, respectively.

The determination of global minimum structures of $(Fe_2O_3)_n$ nanoclusters with $n$ = 6–10 used a two stage procedure. First, for each cluster size global structure optimizations employing GA in combination with the new IP-$Fe_2O_3$ were performed. The central part of the evolutionary algorithm is the 'cut and slice' crossover as well as mutation operator, which acts on every cluster population (*cf.* Ref. 16 for more details). At least two independent GA runs with up to 4000 local optimizations in each case (80 generations, 50 structures per population) were performed yielding identical low energy isomers. Next, the structures of the 50 most stable isomers were refined at the DFT level. Since the geometric structure and relative energies of different isomers of larger $(Fe_2O_3)_n$ clusters are only weakly dependent of the precise magnetization (spin) state[33] the present study assumed ferromagnetic states for $n$ > 6. The DFT calculations were performed using the TURBOMOLE program package[64-66] along with the B3-LYP exchange-correlation functional[67-69] and triple-zeta valence plus polarization (TZVP) basis sets for all atoms.[70] The multipole accelerated resolution of the identity (MARI-J) method[71] for the Coulomb term employing the corresponding auxiliary basis sets[72] was applied to accelerate the calculations. The binding energies $\Delta E_b$ of the clusters with respect to the ground state isomer of $Fe_2O_3$ were computed as energies of the reaction:

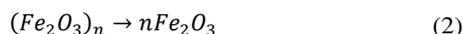

$$(Fe_2O_3)_n \rightarrow nFe_2O_3 \qquad (2)$$

Molecular dynamics simulations of larger $(Fe_2O_3)_n$ NP were performed employing the Large-scale Atomic/Molecular Massively Parallel Simulator (LAMMPS)[73] along with the new IP-$Fe_2O_3$. Initial configurations were constructed as spherical cutouts from α-$Fe_2O_3$ with diameters of 1 ($n$ = 80, **NP1**), 3 ($n$ = 282, **NP3**) and 5 nm ($n$ = 1328, **NP5**). The simulated crystallization used the following procedure. First, the initial structures were equilibrated at 2000 K for 1 ns. Next, for each NP size six different models of the molten NP taken from the second half of the equilibration phase were cooled down from 2000 to 300 K and subsequently optimized. The cooling procedure applied linear velocity scaling for 1 ns in the temperature ranges of 2000-1500 K and 1000-300 K. Between 1500 and 1000 K velocity scaling was applied for 10 ns. All MD simulations were carried out using a time step of 2 fs and without periodic boundary conditions. To verify the independence of the final NP structures from the initial configuration a second 5 nm large $(Fe_2O_3)_n$ NP model was constructed as a spherical cutout from γ-$Fe_2O_3$ and subsequently annealed applying the same procedure (see ESI†).

The determination of the melting temperature of bulk α-$Fe_2O_3$ used a series of independent MD simulations at increasing temperatures between 1800-2200 K with a step of 50 K. At each temperature, a 3D periodic 3×5×2 supercell of α-$Fe_2O_3$ (orthogonal unit cell) was equilibrated for 1 ns employing the isothermal-isobaric (*NPT*) ensemble. The Nosé-Hoover thermostat and barostat were used following the equations of motion of Shinoda *et al.*[74,75] along with a target pressure of 1 atm and a time step of 2 fs. The average potential energy of the system was evaluated for the last 200 ps.

## Experimental methods

### Preparation of $Fe_2O_3$ NP by LAVA

For comparison with the computational results, $Fe_2O_3$ NP were prepared by $CO_2$ laser vaporization. For that purpose a coarse-grained α-$Fe_2O_3$ powder (Aldrich, purity > 99%, grain sizes < 5 μm) was used as the starting material. A $CO_2$ laser beam (wavelength $\lambda$ = 10.59 μm, continuous radiation power $P$ = 2 kW, focus intensity $I$ = 175 kW cm$^{-2}$) was focused onto the raw powder. Absorbing the high-intensity radiation the hematite powder vaporized into plasma. The vaporization proceeded at atmospheric pressure in a continuously flowing aggregation gas (air, total volume flow rate $\dot{V}_{tot}$ = 14.5 m$^3$ h$^{-1}$). The expansion of the plasma into the aggregation gas led to an instant cooling, thereby initiating a rapid condensation and subsequently crystallization of ultrafine $Fe_2O_3$ particles. Finally, the NP were separated from the gas by means of a candle filter. Further details of the LAVA setup and the applied process conditions were described earlier.[12,39] The obtained $Fe_2O_3$ NP were investigated using a high-resolution transmission electron microscope (JEM 3010, JEOL, accelerating voltage $U$ = 30 kV).

## Results and discussion

### *Ab initio* parameterized IP-$Fe_2O_3$

The final parameters of the new IP-$Fe_2O_3$ derived in this work are listed in Table 1. Figure 1 shows the comparison of relative energies for isotropic deformations of α-, γ- and ε-$Fe_2O_3$ unit cells calculated using IP-$Fe_2O_3$ and DFT. The agreement is very good over the whole range of deformations with the root mean square deviation (RMSD) of 4.0, 2.4 and 4.6 kJ/(mol $Fe_2O_3$) for α-, γ- and ε-$Fe_2O_3$, respectively.

Table 1 IP-$Fe_2O_3$ parameters (*cf.* eqn. 1). Ionic charges [e] are $q_{Fe}$ = 1.311 and $q_O$ = -0.874 for Fe and O, respectively.

| Pair $i$-$j$ | $A_{ij}$ [eV] | $\rho_{ij}$ [Å] | $C_{ij}$ [eV Å$^{-6}$] |
|---|---|---|---|
| Fe-O | 62775.704 | 0.165 | 32.055 |
| O-O | 3843.644 | 0.305 | 123.029 |
| Fe-Fe | 2500.943 | 0.029 | 6.383 |

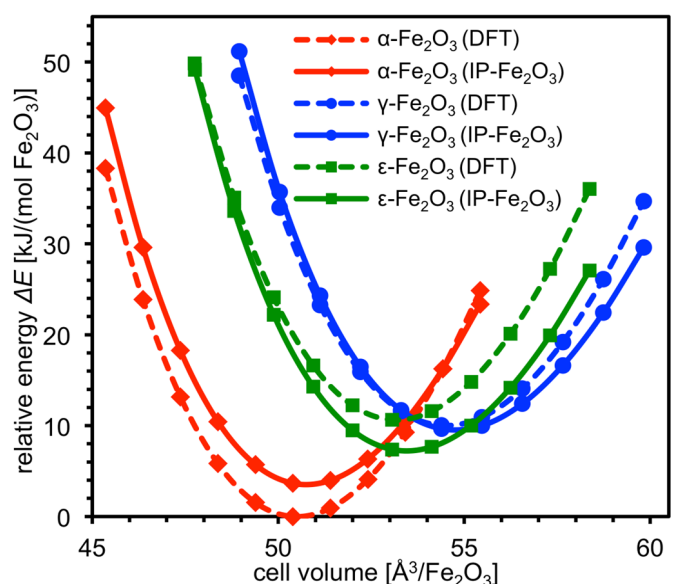

Fig. 1 Relative energies for isotropic deformations of α-, γ- and ε-$Fe_2O_3$ bulk structures calculated using the new *ab initio* parameterized IP-$Fe_2O_3$ and DFT.

Table 2 compares relative stabilities of known $Fe_2O_3$ polymorphs calculated using DFT, IP-$Fe_2O_3$ and other Born-Mayer type interatomic potential functions (BM1-BM3).[76–78] The IP-$Fe_2O_3$ shows the best agreement with the DFT results and gives reasonable stability order of the polymorphs. In contrast to BM1, BM2 and BM3 potentials it predicts α-$Fe_2O_3$ as the most stable phase. It is noteworthy that although β-$Fe_2O_3$ was not included in the training set the IP-$Fe_2O_3$ properly reproduces it as the least stable $Fe_2O_3$ polymorph.

Table 2 Relative stabilities [kJ/(mol $Fe_2O_3$)] of known $Fe_2O_3$ polymorphs with respect to α-$Fe_2O_3$ calculated using DFT and various IP.

|  | β-$Fe_2O_3$ | γ-$Fe_2O_3$ | ε-$Fe_2O_3$ |
|---|---|---|---|
| DFT[a] | 24.80 | 10.03 | 10.64 |
| IP-$Fe_2O_3$[a] | 39.48 | 17.11 | 13.30 |
| BM1[76] | -66.36 | -20.80 | -9.95 |
| BM2[77] | -37.73 | 40.71 | 25.95 |
| BM3[78] | -62.78 | -13.21 | 0.69 |

[a] This study.

Table 3 shows the comparison of DFT and IP-$Fe_2O_3$ calculated cell parameters and bulk moduli for $Fe_2O_3$ polymorphs. For both methods the cell parameters differ less than 1% from the experimental values. Similarly, the calculated bulk moduli of α- and γ-$Fe_2O_3$ show a very good agreement with experimental data. These results demonstrate that IP-$Fe_2O_3$ not only accurately reproduces DFT results but also known experimental data for $Fe_2O_3$ polymorphs.

Table 3 Comparison of cell parameters $a$, $b$, $c$ [Å] and bulk moduli $B$ [GPa] calculated by IP-$Fe_2O_3$ and DFT with experimental data. Deviations of bulk moduli are given with reference to the experimental values in bold face.

| Polymorph | $a$ | $b$ | $c$ | $B$ |
|---|---|---|---|---|
| α-$Fe_2O_3$ (exp)[a] | 5.038 | 5.038 | 13.772 | 178-231, **199** |
| DFT | +0.013 | +0.013 | -0.087 | +5.6 |
| IP-$Fe_2O_3$ | -0.041 | -0.041 | +0.127 | -7.5 |
| β-$Fe_2O_3$ (exp)[b] | 9.404 | 9.404 | 9.404 | - |
| DFT | +0.019 | +0.018 | +0.016 | - |
| IP-$Fe_2O_3$ | +0.004 | +0.004 | +0.004 | - |
| γ-$Fe_2O_3$ (exp)[c] | 8.332 | 8.332 | 25.113 | 203-213, **190** |
| DFT | +0.016 | +0.016 | -0.139 | -0.8 |
| IP-$Fe_2O_3$ | +0.117 | +0.117 | -0.089 | -11.4 |
| ε-$Fe_2O_3$ (exp)[d] | 5.095 | 8.789 | 9.437 | - |
| DFT | +0.007 | -0.002 | +0.032 | - |
| IP-$Fe_2O_3$ | -0.013 | 0.000 | +0.068 | - |

[a] Cell parameters taken from Ref. 45, bulk moduli from Refs 79-83.
[b] Cell parameters taken from Ref. 48.
[c] Cell parameters taken from Ref. 84, bulk moduli from Refs 85-87.
[d] Cell parameters taken from Ref. 47.

MD simulations of bulk α-$Fe_2O_3$ at increasing temperatures employing IP-$Fe_2O_3$ yield a sudden increase of the average potential energy between 2000 and 2050 K. This corresponds to the melting temperature of about 2025 K (see ESI†), in a good agreement with the experimentally observed melting and decomposition point at 1835 K.[1] The difference of 200 K to the experiment is not unexpected since MD simulations use a perfect solid model neglecting the effects of surfaces, material decomposition and lattice defects. Similar overestimation of the melting temperature was also reported for bulk $Al_2O_3$.[88]

## ($Fe_2O_3)_n$ nanoclusters ($n$ = 1-10)

Figure 2 shows tentative global minima of $(Fe_2O_3)_n$ clusters with $n$ = 1-10. It includes also structures with relative energies of less than 1 kJ/(mol $Fe_2O_3$) with respect to the global minimum and at least the two most stable ones (see ESI†). The cluster structures for $n$ = 1–5 are taken from our recent communication[33] and are included here for the sake of completeness.

The two most stable isomers of $Fe_2O_3$ consist of a two-membered Fe–O ring and a terminal O atom. The planar, $C_{2v}$ symmetric **1A** with its $^1B_1$ ground state (GS) is the global minimum. This structure has been reported as the most stable for $Fe_2O_3$[29,30] and $Fe_2O_3^-$.[25] The second most stable isomer is the angled **1B** with the $^3A``$ GS, similar to the most stable configuration of the quartet GS of $Fe_2O_3^+$.[26] The global minimum **2A** of $(Fe_2O_3)_2$ with its antiferromagnetic (AF) $^1A_2$ GS assumes the adamantane-like $C_{2v}$ symmetric structure. The open sheet-like, $C_2$ symmetric **2B** with the $^{11}B$ GS consists of five fused two-membered Fe-O rings and is 30.6 kJ/(mol $Fe_2O_3$) higher in energy. Both **2A** and **2B** have been reported as the most and the second most stable structure of $(Fe_2O_3)_2$, respectively.[31,32] However, these studies predicted either a ferrimagnetic, $C_{3v}$ symmetric[31] or a FM, $T_d$ symmetric $^{21}A_1$ GS[32] for **2A** and a FM, $C_{2h}$ symmetric $^{21}B_g$ GS for **2B**. For $(Fe_2O_3)_2^-$ an AF state of **2A** was reported as the most stable spin configuration.[24]

The global minimum of $(Fe_2O_3)_3$ is the compact, $C_1$ symmetric **3A** with the AF GS. The second most stable $C_s$ symmetric **3B** with the $^1A´$ GS is an open, sheet-like structure containing exclusively two-membered Fe–O rings. It is 15.1 kJ/(mol $Fe_2O_3$) less stable than the global minimum. The two most stable isomers **4A** and **4B** of $(Fe_2O_3)_4$ are compact, $C_1$ symmetric structures with the AF GS and relative energy difference of only 4 kJ/(mol $Fe_2O_3$). For $(Fe_2O_3)_5$ the tower-like **5A** with the AF GS is the global minimum. The second most stable compact **5B** with its ferrimagnetic $^{11}A$ GS is 7.3 kJ/(mol $Fe_2O_3$) higher in energy.

Starting from $n = 6$ all $(Fe_2O_3)_n$ clusters contain the tetrahedral adamantane-like (TAL) structural element similar to **2A**. The global minimum of $(Fe_2O_3)_6$ is the $C_{2h}$ symmetric **6A** build up of a central cage unit fused with two TAL units. The next most stable $C_1$ symmetric, compact **6B** is only 0.4 kJ/(mol $Fe_2O_3$) less stable than **6A**. Similarly, $(Fe_2O_3)_7$ shows two low energy isomers **7A** and **7B** that are separated by only 0.7 kJ/(mol $Fe_2O_3$). The global minimum **7A** of $(Fe_2O_3)_7$ is $C_s$ symmetric and consists mainly of two- and three-membered Fe–O rings, with two of the three-membered rings bridged by one O atom. **7B** is a compact structure with no symmetry elements.

For $(Fe_2O_3)_8$ the global structure optimization procedure yields four isomers with relative energies within less than 1 kJ/(mol $Fe_2O_3$). The $C_2$ symmetric **8A** is the most stable structure. Among the higher energy isomers **8C** exhibits $C_s$ symmetry and resembles the structure of **7A**. **8B** and **8D** have no symmetry elements.

In contrast to $(Fe_2O_3)_8$ the two most stable isomers **9A** and **9B** of $(Fe_2O_3)_9$ are energetically well separated with a relative energy difference of 5.2 kJ/(mol $Fe_2O_3$). The TAL unit can be considered as the main building block of both clusters. **9A** belongs to the symmetry point group $C_{2v}$ and has nearly tetrahedral structure. This structure bears a resemblance to **7A** and **8C**. The next most stable **9B** shows a lower $C_s$ symmetry but very similar structure. In case of $(Fe_2O_3)_{10}$ the global optimization procedure yields four structures with relative energies below 1.0 kJ/(mol $Fe_2O_3$), all consisting mainly of the TAL building units.

As a general feature the most stable $(Fe_2O_3)_n$ clusters assume compact structures with little or no symmetry. One exception is the nearly $T_d$ symmetric TAL unit of **2A**. For $n = 2$-$5$ the clusters contain mainly two and three-membered Fe-O rings. Some isomers, in particular those with no symmetry elements contain also larger Fe-O rings, *e.g.*, **3A**. Starting from $n = 5$ the clusters start to assume increasingly tetrahedral shape with TAL unit as the main building block. However, the small energy differences between different isomers of the same cluster size make precise structural assignment for larger $(Fe_2O_3)_n$ clusters difficult.

Figure 3 summarizes the main IP-$Fe_2O_3$ and DFT results for $(Fe_2O_3)_n$ clusters with $n = 2$–$10$: mean Fe–O bond lengths, binding energies of global minimum structures and relative stability of the two lowest energy isomers (*n*A and *n*B, *cf.* Fig. 2) for each cluster size. For the sake of completeness DFT results for the ground state and FM state of $(Fe_2O_3)_n$ clusters with $n = 2$–$5$ are included taken from our recent communication.[33]

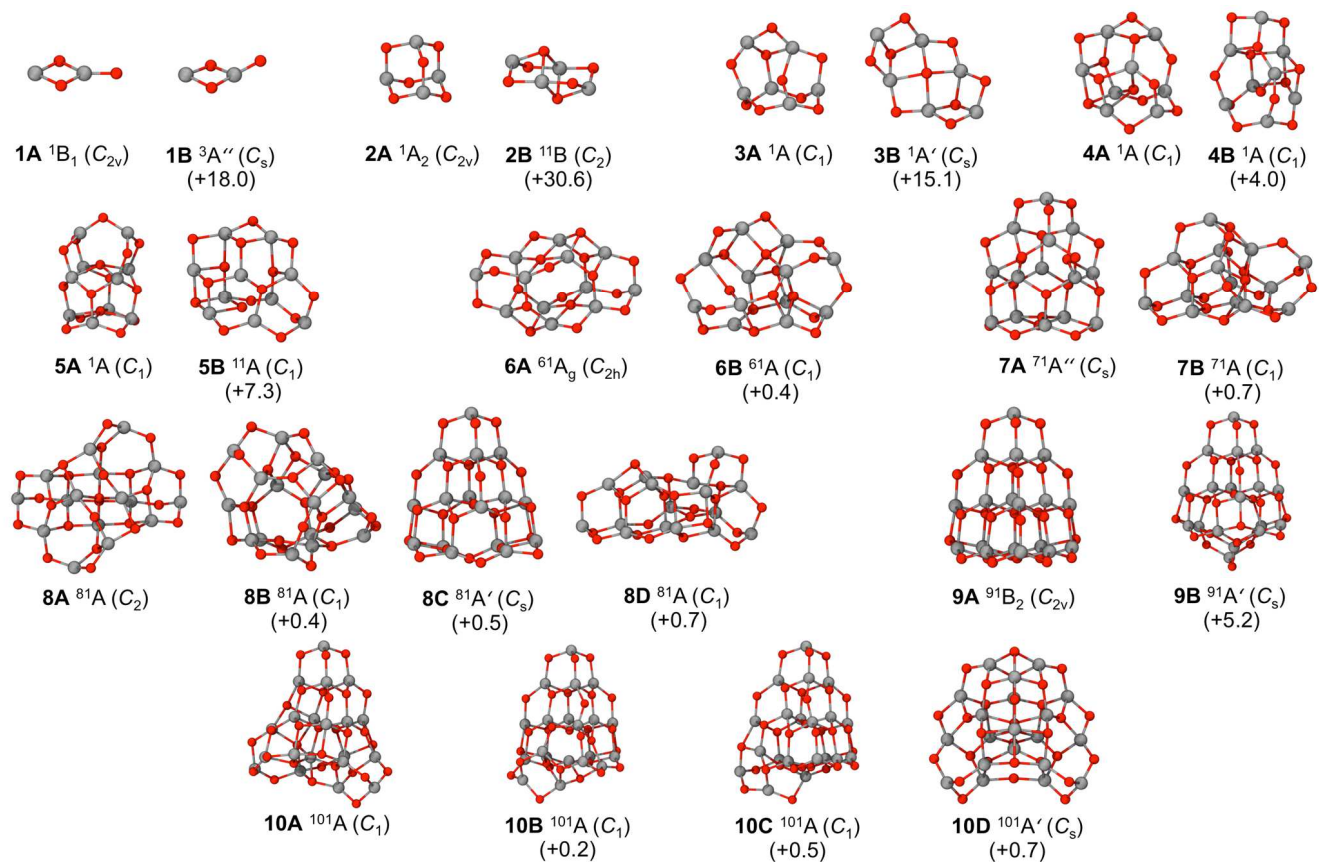

Fig. 2 Low-energy structures of (Fe$_2$O$_3$)$_n$ clusters with $n$ = 1–10. Relative energies with respect to the global minimum are given in parentheses [kJ/(mol Fe$_2$O$_3$)]. Fe grey, O red.

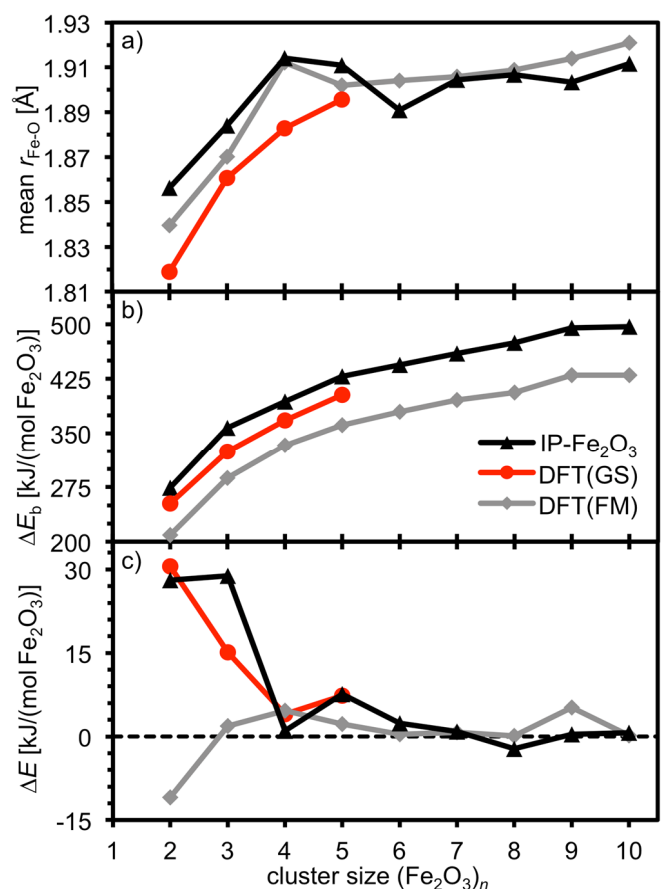

Fig. 3 Comparison of IP-Fe$_2$O$_3$ and DFT results for (Fe$_2$O$_3$)$_n$ clusters with $n$ = 2–10: (a) average Fe–O bond lengths, $r_{Fe-O}$, (b) binding energies, $\Delta E_b$, of the most stable clusters and (c) relative stability, $\Delta E$, of the two lowest energy isomers ($n$A and $n$B, *cf.* Fig. 2). GS and FM indicate ground and ferromagnetic states, respectively.

The transferability and reliability of the IP-Fe$_2$O$_3$ is demonstrated by the very good agreement of structural parameters with those obtained at the DFT level. For (Fe$_2$O$_3$)$_n$ clusters with $n$ = 2–5 the mean Fe–O bond lengths deviate less than 0.05 Å (Fig. 3a) between IP-Fe$_2$O$_3$ and DFT for GS as well as FM states. As already mentioned this shows that the precise magnetization (spin) state of the clusters has only small influence on their geometric structure.[33] The agreement between DFT and IP-Fe$_2$O$_3$ optimized structures is even better for clusters with $n$ = 6–10, with the mean Fe–O bond lengths deviation of less than 0.02 Å. In addition, the general trend of increasing Fe–O bond distances with increasing cluster size is very well reproduced by IP-Fe$_2$O$_3$.

Figure 3b shows the cluster size dependence of binding energy, $\Delta E_b$, for the lowest-energy structures. For (Fe$_2$O$_3$)$_n$ clusters with $n$ = 2–5 the difference between the binding energies of the GS and FM state is equal to the relative energy of the corresponding spin configurations (*cf.* Ref. 33). It shows only small variation with the cluster size. The IP-Fe$_2$O$_3$ results are in a good agreement with the DFT values and show a constant shift of about 75 kJ/(mol Fe$_2$O$_3$) towards lower binding energies. IP-Fe$_2$O$_3$ also properly reproduces the monotonic increase of $\Delta E_b$ with increasing cluster size.

Figure 3c compares the relative energies of the two most stable (Fe$_2$O$_3$)$_n$ isomers calculated using IP-Fe$_2$O$_3$ and DFT. The agreement between both methods is very good, within 5 kJ/(mol Fe$_2$O$_3$), over the whole range of cluster sizes, with exception of $n$ = 3 where the deviation of about 15 kJ/(mol Fe$_2$O$_3$) is somewhat higher. A larger deviation for such small clusters is not unexpected since the training set of IP-Fe$_2$O$_3$ contains only bulk structures. The small energy difference between the GS and FM state for larger clusters show that relative stabilities of larger (Fe$_2$O$_3$)$_n$ clusters are virtually independent of the precise magnetization state.[33] This supports our approach for determination of low-energy cluster structures employing GA in combination with IP-Fe$_2$O$_3$ followed by structure refinement at the DFT level assuming FM states.

Our findings demonstrate that IP-Fe$_2$O$_3$ can accurately describe the structures and relative stabilities of both small (Fe$_2$O$_3$)$_n$ clusters and bulk Fe$_2$O$_3$. Therefore, one can expect that IP-Fe$_2$O$_3$ is also well suited for simulations of larger (Fe$_2$O$_3$)$_n$ NP at an intermediate length scale between clusters and bulk material.

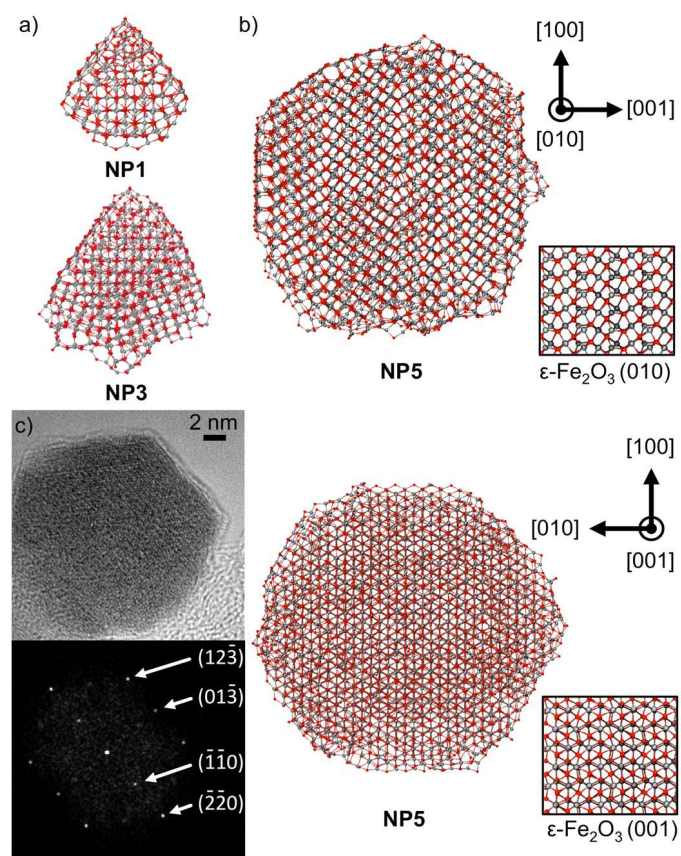

Fig. 4 Structure of (Fe$_2$O$_3$)$_n$ nanoparticles: (a) **NP1** ($n$ = 80) and **NP3** ($n$ = 282), (b) comparison of **NP5** ($n$ = 1328) with two different lattice planes of ε-Fe$_2$O$_3$ and (c) high-resolution TEM micrograph of a LAVA synthesized ε-Fe$_2$O$_3$ NP and its indexed Fourier transform (ε-Fe$_2$O$_3$, zonal axis [$\bar{3}\bar{3}1$]). The coordinate systems refer to the NP orientation. Fe grey, O red.

## $(Fe_2O_3)_n$ nanoparticles ($n$ = 80 - 1328)

The most stable $(Fe_2O_3)_n$ structures with $n$ = 80, 282 and 1328 determined by the simulated crystallization procedure are depicted in Figure 4. For **NP5**, two orientations are shown along with two lattice planes of ε-$Fe_2O_3$ (Fig. 4b). Figure 4c shows a high-resolution transmission electron microscopy (TEM) micrograph of a 18 nm $Fe_2O_3$ NP prepared by the LAVA process. Details of the simulated crystallization of **NP5** including temperature dependence of potential energy and coordination number (CN) distributions of Fe and O atoms are given in Figures 5a and 5b, respectively. To facilitate comparison with bulk $Fe_2O_3$ polymorphs and discern structure differences between the inner and surface parts of **NP5** the CN distribution is evaluated separately for the core part (4 nm in diameter) and the whole **NP5**, denoted as $CN_c$ and $CN_e$, respectively.

The most stable configurations of $Fe_2O_3$ NP with 1 (**NP1**) and 3 nm (**NP3**) diameter show tetrahedral, wedge-like morphology (Fig. 4a) also present in smaller $(Fe_2O_3)_n$ clusters, such as **7A**, **8C**, **9A**, **9B** and **10A** (*cf.* Fig. 2). This particular shape can be rationalized by the presence of a large number of the TAL $(Fe_2O_3)_2$ building blocks forming a surface layer and a high surface-to-volume ratio of **NP1** and **NP3**. This is different in case of **NP5** that consists of a single crystalline, hexagonally shaped domain. In order to investigate the reproducibility of the crystallization process of **NP5** we repeated the procedure for several independent initial configurations. In all cases similar NP structures were obtained that are in a narrow energy window, less than 9 kJ/(mol $Fe_2O_3$) compared to **NP5**. We have also used an initial configuration constructed as a spherical cut of γ-$Fe_2O_3$. It resulted in a structure virtually identical to **NP5** (see ESI†).

During the phase transition from the liquid to solid state the corresponding latent heat is released from the system. This can be seen as a sudden drop of potential energy of **NP5** between 1235 and 1215 K (Fig. 5a), yielding the melting temperature of about 1225 K. This value is significantly lower than the melting and decomposition point of bulk $Fe_2O_3$ (1835 K). This size-dependent melting point depression of nanoparticulate materials connected to their large specific surface area is a well-know phenomenon.[89] Using the calculated α-$Fe_2O_3$ melting temperature of 2025 K (see ESI†) the melting point depression for **NP5** is about 800 K. This value is probably somewhat overestimated due to high cooling and heating rates of **NP5** and bulk α-$Fe_2O_3$, respectively, during our MD simulations.

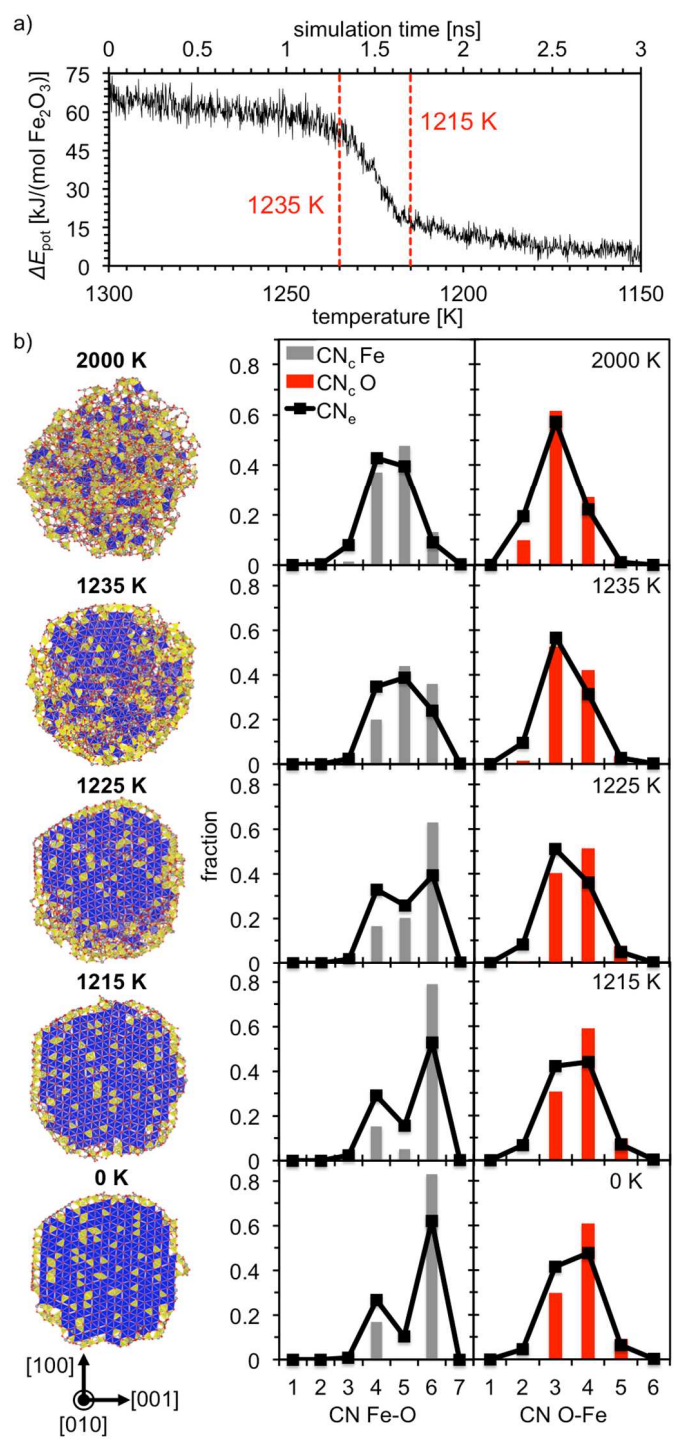

Fig. 5 MD simulation of the crystallization process of NP5: (a) change of the potential energy ($\Delta E_{pot}$) at the melting point and (b) coordination number (CN) distribution of Fe and O atoms, respectively, as a function of temperature. CN distributions are calculated for the core ($CN_c$) and the entire NP5 ($CN_e$). For each temperature cross-sections ((010) plane of the final ε-$Fe_2O_3$ crystallite) are shown highlighting Fe atoms with octahedral (blue) and tetrahedral (yellow) coordination.

The structural evolution of **NP5** during crystallization is shown in Figure 5b along with its cross-sections along [010] direction of the final ε-$Fe_2O_3$ crystal and temperature dependence of the coordination number (CN) distribution for Fe and O atoms. In general, $CN_e$ is shifted towards lower values due to a larger number of low-coordinated surface atoms. Equilibration of the initial spherical cuts of bulk $Fe_2O_3$ for 1 ns at 2000 K is sufficient to generate melted configurations that are independent of the initial structure (see ESI†). At this temperature, the CN distribution shows mainly four- and five-fold coordinated Fe as well as three-fold coordinated O atoms. Decreasing temperature increases the mean $CN_e$ from 3.9 and 3.0 at 2000 K to 5.3 and 3.6 at 0 K for Fe and O atoms, respectively. At 1235 K the corresponding NP cross-sections indicate beginning of crystallization with a nucleation center forming close to the NP surface. Compared to 2000 K both $CN_e$ and $CN_c$ distributions show an increased fraction of six-fold coordinated Fe and four-fold coordinated O atoms. In the temperature range from 1235 to 1225 K advancing crystallization transforms five-fold coordinated Fe atoms to six-fold coordinated ones and three-fold coordinated O atoms into four-fold coordinated ones. For temperatures between 1225 and 1215 K CN distributions show a significant reduction of five-fold coordinated Fe and emergence of five-fold coordinated O atoms. In contrast, the fraction of four-fold coordinated Fe atoms remains virtually constant. The CN histograms for temperatures below 1215 K show no significant structural changes indicating complete crystallization. The somewhat irregular distribution of four-fold coordinated Fe atoms seen in the cross-section at 0 K arises from the presence of defect sites within the NP core.

Comparison of structural characteristics of the final **NP5** with bulk $Fe_2O_3$ polymorphs indicates that it consists of a single, albeit imperfect ε-$Fe_2O_3$ crystal. Analysis of $CN_c$ yields 0.17 and 0.83 as fractions of Fe atoms with tetrahedral and octahedral coordination, respectively (*cf.* Fig. 5b, 0 K). These values are close to the fractions of four- (0.25) and six-fold (0.75) coordinated Fe atoms in bulk ε-$Fe_2O_3$. In contrast, γ-$Fe_2O_3$ contains a considerably higher fraction (0.375) of Fe atoms in tetrahedral coordination and α- as well as β-$Fe_2O_3$ contain exclusively six-fold coordinated iron. Furthermore, $CN_c$ distribution shows presence of five-fold coordinated O atoms which are a unique structural feature of ε-$Fe_2O_3$. Deviations from the ideal CN distribution of ε-$Fe_2O_3$ are related to lattice defects present in **NP5** such as vacancies and dislocations. Similar deviations of CN fractions due to lattice disorders within nanoparticulate ε-$Fe_2O_3$ were also reported in experimental studies.[47]

The final confirmation comes from the comparison of **NP5** with $Fe_2O_3$ nanoparticles synthesized in the LAVA process. Figure 4c displays a high-resolution TEM micrograph of such an NP with a diameter of about 18 nm showing the typical hexagonal morphology as well as the corresponding Fourier transform (FT). The lattice planes are visible throughout the whole particle, suggesting a single crystalline structure although the presence of defects cannot be ruled out. The FT of the lattice fringes was indexed according to the structural model of Tronc *et al.*[47] and identifies the NP as ε-$Fe_2O_3$ aligned along its [$\bar{3}31$] axis. Our earlier study shows that larger $Fe_2O_3$ NP with diameters in the range of 50 nm can also display octagonal morphology.[12] The hexagonal shape of the LAVA synthesized $Fe_2O_3$ NP is very well reproduced by the results of simulated crystallization of **NP5** (Fig. 4b).

Our earlier study of LAVA synthesized $Fe_2O_3$ nanopowders reported formation of different ratios of ε-$Fe_2O_3$ and γ-$Fe_2O_3$, depending on oxygen concentration in the condensation atmosphere.[12] The formation of the two phases is connected to their very similar stability (*cf.* Table 2) and was attributed to differences in the nucleation kinetics due to the presence of remarkably stable iron-ozone complexes. They act as precursors for octahedrally coordinated Fe sites upon rapid condensation and solidification of nanoparticles. Therefore, higher concentration of ozone in oxygen-rich condensation gas leads to an increased amount of ε-$Fe_2O_3$ with a larger number of octahedrally coordinated Fe atoms. In contrast, oxygen-poor atmosphere results in formation of a higher fraction of γ-$Fe_2O_3$ containing significantly less octahedral Fe sites. Strong dependence of ε-$Fe_2O_3$ content on experimental condition was also reported for other synthetic routes.[6] The appearance of only ε-$Fe_2O_3$ phase during simulated crystallization of **NP5** can be rationalized by the absence of kinetic factors influencing the nucleation process. All simulated crystallizations were performed for stoichiometric $Fe_2O_3$ systems starting from a well equilibrated, molten state. In contrast, NP formation by the rapid condensation of LAVA generated plasma is a non-equilibrium process that involves different non-stoichiometric gas phase species present during nucleation. This can result in formation of thermodynamically metastable phases.[12] However, in the present case the very good agreement between results of simulated crystallization of **NP5** and structure analysis of LAVA generated $Fe_2O_3$ NP provides the first direct indication that ε-$Fe_2O_3$ may be thermodynamically the most stable phase in this size range. The significant melting point depression of 800 K found in this study provides explanation for thermal instability of small ε-$Fe_2O_3$ NP observed by several authors[6] due to sintering and formation of larger agglomerates favoring conversion to α-$Fe_2O_3$. Indeed, it has been reported[6] that restricting growth of $Fe_2O_3$ NP by isolation in a $SiO_2$ matrix[90] or special synthesis conditions[8] significantly enhances stability of the ε-$Fe_2O_3$ phase.

## Conclusions

In summary, atomic structure and properties of nanoparticulate $Fe_2O_3$ are characterized starting from its basic $Fe_2O_3$ building unit and $(Fe_2O_3)_n$ nanoclusters through to nanometer-sized $Fe_2O_3$ particles. This has been achieved combining global structure optimizations of nanoclusters, simulated crystallizations of lager $Fe_2O_3$ NP and laser vaporization experiments. The key computational tool are carefully parameterized, *ab initio* derived interatomic potential functions. With exception of nearly tetrahedral, adamantane-like $Fe_2O_3$ dimer $(Fe_2O_3)_n$ nanoclusters assume compact, virtually amorphous structures with little or no symmetry. For $n$ = 2–5 they consist of mainly two and three-membered Fe–O rings. Starting from $n$ = 5 they increasingly assume tetrahedral shape with adamantane-like $(Fe_2O_3)_2$ unit as the main building block. In case of larger nanoclusters small energy differences between different isomers make precise structural assignment difficult. The tetrahedral morphology persists for Fe$_2$O$_3$ NP with diameters up to 3 nm. Simulated crystallization of larger NP with diameters of about 5 nm demonstrates pronounced melting point depression of 800 K and leads to formation of ε-Fe$_2$O$_3$ single crystals with hexagonal morphology. This is in a very good agreement with the results of structure analysis of LAVA generated Fe$_2$O$_3$ nanopowders providing the first direct indication that ε-Fe$_2$O$_3$ may be thermodynamically the most stable phase in this size range. The observed significant melting point depression provides explanation for thermal instability of small ε-Fe$_2$O$_3$ NP due to sintering and formation of larger agglomerates facilitating conversion to α-Fe$_2$O$_3$.

## Acknowledgements

Authors gratefully acknowledge financial support from the Ministry of Education, Science and Culture of the state Thuringia, Fonds der Chemischen Industrie and Turbomole GmbH.

## Notes

* Otto Schott Institute of Materials Research, Friedrich Schiller University of Jena, Löbdergraben 32, 07743 Jena, Germany. Email: marek.sierka@uni-jena.de

† Electronic Supplementary Information (ESI) available: See DOI: 10.1039/b000000x/